# On the Reliability of Seasonal Climate Forecasts

by

A. Weisheimer and T.N. Palmer

National Centre for Atmospheric Science (NCAS), Department of Physics, University of Oxford, UK

and

European Centre for Medium-Range Weather Forecasts (ECMWF), Reading, UK

**Abstract**

Seasonal climate forecasts are being used increasingly across a range of application sectors. A recent UK governmental report asked: How good are seasonal climate forecasts on a scale of 1-5 (where 5 is very good), and how good can we expect them to be in 30 years time? Seasonal climate forecasts are made from ensembles of integrations of numerical models of climate. We argue that "goodness" should be assessed primarily in terms of the probabilistic reliability of these ensemble-based forecasts and that a `5' should be reserved for systems which are not only reliable overall, but where, in particular, small ensemble spread is a reliable indicator of low ensemble forecast error. We study the reliability of regional temperature and precipitation forecasts of the current operational seasonal forecast system of the European Centre for Medium-Range Weather Forecasts, universally regarded as a world leading operational institute producing seasonal climate forecasts. A wide range of "goodness" rankings, depending on region and variable (with summer forecasts of rainfall over Northern Europe performing exceptionally poorly) is found. Finally, we discuss the prospects of reaching "5" across all regions and variables in 30 years time.

1. **Introduction**

Over the last 30 years, the science of predicting seasonal timescale fluctuations in weather has grown from a research activity undertaken in a few academic and research institutes (*eg Cane et al, 1986*), to a routine operational activity in a number of meteorological forecast services (*Arribas et al., 2011*; *Stockdale et al., 2011*; *Saha et al., 2013).* Unlike conventional weather forecasts, seasonal predictions do not attempt to forecast the detailed day-to-day evolution of weather. Such detailed prediction is ruled out by the chaotic nature of the climate system, otherwise known as the "Butterfly Effect" (*Lorenz, 1963*). Rather, seasonal predictions provide estimates of seasonal-mean statistics of weather, typically up to three months ahead of the season in question. Hence, for example, a seasonal forecast can provide information on how likely it is that the coming season will be wetter, drier, warmer, or colder than normal. The physical basis for such estimates arises from the effect of predictable seasonal-timescale signals arising from the ocean, and to a lesser extent the land surface, on the atmosphere (*Palmer and Anderson, 1994*). The key paradigm for seasonal forecasting is El Niño, a coupled ocean-atmosphere phenomenon occurring primarily in the tropical Pacific and predictable six months and more ahead (*Jin et al., 2008; Weisheimer et al., 2009*).

Such information is relevant to a variety of users in weather-sensitive sectors, and therefore can influence decisions made in these sectors. As a result, seasonal climate forecasts are increasingly being used across a range of application areas; see *Dessai and Soares (2013)* for a recent review. For example, information about seasonal average rainfall and temperature for the growing season can potentially influence a farmer's decision about which crops to plant ahead of time, or a humanitarian organisation's strategy for anticipating food shortages in drought-prone regions of the developing world. However, this information is only useful if it is skilful.

In the literature there exists a plethora of methods to estimate the skill of forecasts (*Jolliffe and Stephenson, 2011*). In general, each of these methods quantifies a different detailed aspect of the forecast quality. In this paper, however, we try to simplify the question. Rather than coming up with complex estimates of different characteristics of forecast skill whose relevance strongly depends on the specific application, we simply ask: On a scale of 1-5, where 5 is very good, how skilful are seasonal forecasts today? On a similar scale, how skilful can we expect seasonal forecasts to be 30 years from now? These types of question are sufficiently open-ended that they may appear impossibly difficult to answer in any such succinct way. And yet precisely these types of question are being asked by policy makers e.g. by the UK Government as it considers options for future investment in science (*Foresight, 2012*).

What level of skill should a seasonal forecast system reach, to merit being rated a "5"? Would every forecast of seasonal-mean temperature and rainfall for a particular region have to be both precise and accurate? This is likely to be an impossible goal. Whilst the impact of the Butterfly Effect is mitigated substantially by focussing on prediction of seasonal-mean rather than instantaneous variables, it is not eliminated entirely. In addition, the coupled ocean-atmosphere models used to make seasonal forecasts are finite truncations of the underlying (partial differential) equations of the climate system, and hence are only approximate representations of reality. These two facts of the matter imply that seasonal predictions must necessarily be considered probabilistic in character – forecasts from any deterministic seasonal prediction will necessarily be unreliable and therefore untrustworthy.

However, this does not imply that any probabilistic forecast system is necessarily reliable. In this paper we define a forecast as being reliable if it demonstrates statistical reliability in the following sense. Consider a set of predictions derived from ensemble forecasts[*]. For some of these cases it is predicted that the chance of above average seasonal-mean rainfall for the coming growing season will be 80%. If the probabilistic forecast system is reliable, then one can expect that in 80% of these predictions the actual seasonal-mean rainfall will be above average. In this way, the concept of "reliability" can be extended to probabilistic forecast systems (*Murphy, 1973*). If a forecast system is unreliable in this sense, then poor decisions can be made. As discussed in Section 2, a farmer might decide that it makes economic sense to grow a particular type of crop when the forecast for above-average rain exceeds 80%. However, if in reality above-average rain only occurs 50% of the time when the forecast probability exceeds 80%, the potential economic benefit of planting the particular crop may be completely lost by the unreliable probabilistic forecast.

This raises an important conceptual point. Consider a hypothetical forecast system which most of the time forecasts climatological probabilities, but occasionally forecasts probabilities which are substantially different from climatology. If the probability forecasts have reliability and if the system can successfully discriminate between predictable and unpredictable situations, we would rate such a system as "5" even though the formal skill scores such as the Brier Skill Score (see below) for such a system may be small. We can compare this with an ensemble forecast system where the forecast probabilities are comparable with climatology for all initial conditions. Here we would rate such a system as "2" – such a forecast system would never lead decision makers to make poor decisions, though it might not be particularly useful.

In this paper we develop objective criteria for classifying forecast skill into five categories, discuss how close we are to achieving a "5" today, and consider what is needed to achieve a "5" in 30 years time. All results presented here are based on the (state of the art) operational seasonal forecast System 4 from the European Centre for Medium-Range Weather Forecasts (ECMWF). Depending on the region and variable being studied, we find examples of all five of our categories.

---

[*] Each member of the ensemble uses slightly different initial conditions and different realisations of stochastic representations of sub-grid physical processes in the atmosphere.

## 2. Probabilistic Skill and Decision Making

Forecasts are used to make decisions. For example, a farmer wants to decide what type of crop to plant in the coming season. Suppose there is a choice between two types of crop: *A* and *B*. The crop yield (tons/hectare) $C_A$ and $C_B$ of *A* and *B* depends on a number of meteorological variables such as temperature and precipitation, collectively labelled by *X*. Hence $C_A = C_A(X)$ and $C_B = C_B(X)$. Suppose we have a forecast system which predicts over a given season a probability distribution $\rho(X)$ for *X*. Then the expected crop yield for *A* and *B* is

$$\langle C_A \rangle = \int_X C_A(X)\rho(X)dX$$
$$\langle C_B \rangle = \int_X C_B(X)\rho(X)dX$$

If $\langle C_A \rangle > \langle C_B \rangle$ the farmer might choose *A* over *B*, and vice versa. In practice of course there will be many factors other than climate which determine the farmer's decision, but let us suppose here that climate is the only relevant one.

In general, one can expect $C_A$ to be a nonlinear function of *X*. Hence $\langle C_A \rangle$ will depend on more than just the mode of the distribution $\rho$. The uncertainty, given by the spread of the forecast distribution, might have just as large an impact on the estimate $\langle C_A \rangle$ as does the mode of the forecast distribution.

Let us assume that

$$\langle C_A \rangle_C \equiv \int_X C_A(X)\rho_C(X)dX > \langle C_B \rangle_C \equiv \int_X C_B(X)\rho_C(X)dX$$

where $\rho_C(X)$ is the climatological distribution of *X*. Let us also suppose that in the majority of forecast occasions, the forecast distribution $\rho(X)$ is not significantly different from the climatological distribution $\rho_C(X)$. Then, on these occasions, whilst the farmer is not going to gain any specially useful information from the forecast system, (s)he is not going to be mislead by unreliable information. Conversely, consider the relatively infrequent occasions where $\rho(X) \neq \rho_C(X)$ such that $\langle C_A \rangle < \langle C_B \rangle$. If as a result the farmer decides to grow *B* over *A*, it is essential that the forecast probability function $\rho(X)$ must be reliable.

One way to assess whether such forecast distributions $\rho$ are reliable when $\rho \neq \rho_C$ is to study so-called Attributes (or "Reliability") Diagrams. Reliability diagrams are discussed and shown in Sections 4 and 5. The focus if this paper is the reliability of user-relevant forecast variables in ECMWF's System 4 seasonal forecasts in the situations where $\rho \neq \rho_C$.

### 3. The ECMWF Seasonal Forecast System 4

The European Centre for Medium-Range Weather Forecasts (ECMWF) has been at the forefront of seasonal predictions for many years. Research on predictability on seasonal time scale in the early 1990s (eg. *Palmer and Anderson, 1994*) led to the implementation of the first ECMWF seasonal forecast system based on a global ocean-atmosphere coupled model in 1997, and a successful forecast of the major 1997-98 El Niño (*Stockdale et al., 1998*). This first coupled System 1 was replaced by System 2 in 2001 and System 3 in March 2007. In November 2011 the latest seasonal forecasting System 4 started producing operational forecasts. The results presented in this paper are based on System 4's retrospective seasonal forecasts of 2m temperature and precipitation over land.

The forecasting model of System 4 (*Molteni et al., 2011*) consists of an atmospheric and an oceanic component to simulate the evolution of the global circulation in the atmosphere and in the oceans, based on the physical laws of fluid dynamics. The equations of motions and the thermodynamic laws are solved numerically by discretising the atmosphere and the oceans into several vertical layers and horizontal grid boxes. The atmospheric model component of System 4 is version CY36R4 of ECMWF's weather forecasting model IFS (Integrated Forecasting System). Whereas the model for ensemble weather forecasting is run with a horizontal resolution of approx. 30 km-sized grid boxes, the resolution used in seasonal forecasting of approx. 80 km (spectral resolution of T255) is somewhat coarser. The atmospheric model has 91 vertical levels reaching up to 0.01 hPa. The ocean model used in System 4 is NEMO (Nucleus for European Modelling of the Ocean) version 3.0., a state-of-the-art modelling framework for oceanic research. The ocean model has 42 levels in the vertical and the grid boxes have an approx. length of 110 km (1°).

As discussed in the Introduction, seasonal forecasts must be probabilistic by nature. In practice, probabilistic forecasts are derived by running an ensemble of integrations of the forecast model. Each member of the ensemble uses slightly different initial conditions and different realisations of stochastic representations of sub-grid physical processes in the atmosphere. Operational global forecasts with System 4 are produced at the beginning of each month for forecast lead times of 7 months into the future using 51 ensemble members.

How can we estimate the reliability of these seasonal forecasts? A single probabilistic forecast cannot, in general, be verified or falsified. But for a set of probabilistic forecasts we can evaluate the performance of the forecasting system by systematically comparing the forecasts with observations and by deriving statistical skill measures. These skill estimates, based on the performance of the system in the past, may guide users about the expected performance of the forecasts of the future.

In order to achieve a robust estimate of the System 4 model performance, an extensive set of retrospective forecasts (re-forecasts) of the past has been generated. This forms the basis of the following analysis. The System 4 re-forecasts were started every calendar month over the 30-year period 1981 - 2010 by emulating real forecast conditions when no observed information about the future is available at the beginning of the forecast. Here we analyse (51 member) ensemble re-forecasts initialised on the 1st of May and 1st of November 1981-2010. The forecast lead time is 2-4 months corresponding to the boreal summer (June, July and August, JJA) and winter (December, January and February, DJF) seasons.

In this study, we concentrate our analyses on 2m temperature and precipitation over 21 standard land regions (*Georgi and Francisco, 2000*). The verification data used are ECMWF re-analysis data (ERA-Interim) for 2m temperature (*Dee et al., 2011*) and GPCP for precipitation (*Adler et al., 2003*). As discussed in the Introduction, in seasonal forecasting one is mostly interested in seasonal deviations from the long-term climatological mean. Observed anomalies for each year and season are defined as deviations from the mean over the 1981 to 2010 re-forecast period. In a similar way, model anomalies for each ensemble member were derived from the re-forecasts and the model mean over the re-forecast period. In order to emulate real-time forecast situations, both observed and model anomalies are computed in cross-validation mode by leaving out the actual forecast year in the estimate of the climatological mean values. Transforming absolute temperature and precipitation forecast values into anomalies implicitly also removes any systematic errors, or biases, which develop during the forecasts between the model and the verification.

In the following we will consider dichotomous, or binary, events $E$ based on terciles of the climatological distribution of seasonal anomalies of temperature and precipitation. If $E$ is defined as falling into the lower third of the long-term distribution, the event is called "cold" for 2m temperature or "dry" for precipitation. Similarly, if $E$ relates to the upper third of the distribution, the event is called "warm" or "wet". The seasonal forecasts from System 4 then specify the probability of event $E$ that the seasonal-mean forecast anomalies lie below the lower tercile or above the upper tercile, respectively. Our aim here is to quantify the reliability of such probabilistic tercile event by comparing the forecast probability for $E$ with the corresponding observed frequency of $E$ of the verifying observations.

## 4. Reliability Diagrams and Categories of Reliability

Reliability (or Attributes) Diagrams are tools to display and quantify the statistical reliability of a forecasting system, as defined in the Introduction. Such a diagram graphically summarises for a given binary event $E$ the correspondence of the *forecast* probabilities with the *observed* frequency of occurrence of the event $E$ given the forecast. Reliability is high if this correspondence is very good. Reliability is poor if there is little, no or even negative correspondence between the forecast probabilities and the observed frequencies.

For example, suppose the seasonal forecast probability for event $E$ is equal to 0.8. Then, in a reliable seasonal forecast system, $E$ would actually occur, taking into account sampling uncertainty, on approximately 80% of the cases where $E$ was predicted with a probability of 0.8. A reliability diagram displays a range of such forecast probabilities for *E* and their corresponding observed frequencies collected over the re-forecast period. If the correspondence between the forecast probabilities and the observational frequencies were perfect (and neglecting sampling uncertainty), all data points would lay on a straight diagonal line in the reliability diagram. It is important to note that a forecasting system which always issues the underlying long-term climatological probability of the event has perfect reliability even though it might not provide any additional information to the climatological one.

Figure 1 shows a schematic of a reliability diagram for tercile events *E* but without any data points. Here, the climatological forecast probability and long-term frequency of E in the verification data (by definition 1/3) are denoted by the vertical and horizontal lines. The grey areas in the diagram are linked to forecast situations where the Brier Skill Score† (BSS), based on a no-skill climatological reference, is positive. What does this mean? Often it is of interest to compare seasonal forecasts generated by physical models of the climate system with a reference forecast that serves as a benchmark for the climate model forecasts. Such a comparison allows the definition of skill scores of the forecasts: A skill score is positive (negative) if the forecast is better (worse) than the reference forecast. A widely used reference forecast is the simple climatological long-term mean forecast. For tercile events, the reference forecast would always be 1/3. It can be shown (*Mason, 2004*) that forecast probabilities which fall into the grey indicated areas in Fig 1 contribute positively to the BSS if climatology is used as a reference. The line that separates the skilful from the unskilful forecast probabilities (no-skill line) is defined by BSS=0 indicating that the forecasts below this line are not better than the reference forecast. The no-skill line has, by definition, a slope of 0.5

For any real forecasting system, the data points in a reliability diagram are not likely to lie on a straight line. In order to quantify the overall reliability of an event *E*, and to try to minimise the effects of relatively small statistical samples in estimating reliability, we apply a weighted linear regression as a best-fit estimate on all data points in the reliability diagram using the number of forecasts in each probability bin as weights. The slope of the so derived reliability line can be used as a quantitative measure of the reliability of the system: A slope of ~1 indicates very good reliability; a slope of ~0 a very poor or no reliability. A slope which is negative could be characterised as "worse than useless" as it might encourage decision makers to make decisions which could turn out to be exceptionally poor. It is this slope of the reliability line on which our proposed categories of reliability will be based.

In addition to the best-guess reliability slope we estimate the uncertainty around that slope. Using a bootstrap algorithm with replacement we draw randomly from the set of System 4 re-forecast data and compute the slope of the reliability regression line. By repeating this procedure 1000 times we construct a re-sampled data set of regression line slopes and use the 75 per cent confidence interval of the re-sampling distribution to define an uncertainty range around our best-guess reliability slope.

---

† The BSS is based on the Brier Score (*Brier, 1950*) which can be considered the probabilistic generalisation of the mean squared error for dichotomous events.

In order to answer the question posed in the Introduction - how "good" on a scale of 1 to 5 are our current seasonal forecasts - we propose a categorisation of reliability based on the slope of the reliability line and the uncertainty associated with it. In Figure 2 we show schematics for each of the five categories to demonstrate their definitions; Figure 3 has examples for each category from the System 4 re-forecasts data. Here the size of the data points is proportional to the number of forecasts in that forecast probability bin.

The highest *Category 5* classifies perfect reliability conditions (see Fig. 2a). It is defined such that the uncertainty range of the reliability slope includes the perfect reliability slope of 1 and falls completely into the skilful BSS area. Thus, given the sampling uncertainty, such forecasts are perfectly reliable. Forecasts in category 5 can potentially be very useful for decision making. In Fig 3a we show as an example for category 5 forecasts the reliability diagram for the tercile event of warm DJF over the Sahel region of System 4. Here, the best-guess reliability line is only slightly steeper than the diagonal. The uncertainty range clearly includes the perfect reliability slope of 1 (diagonal).

The second highest *Category 4* is characterised by reliability diagrams where the uncertainty range of the reliability line lies entirely within the skilful area but does not include the perfect reliability line, see schematic in Fig. 2b. It describes forecast reliability that is still very useful for decision making. An example from System 4 is given in Fig. 3b for wet conditions in JJA over East Africa.

If the slope of the reliability line including its uncertainty range is positive but not skilful (reaches below the no-skill line), the forecasts are classified as *Category 3* reliable, see Fig. 2c. Such forecasts can be considered marginally useful for decision making. As an example for System 4, dry DJF forecasts over West Africa as shown in Fig. 3c fall into Category 3 reliability.

If the slope of the reliability line cannot be distinguished, within its uncertainties, from zero, the forecasts are defined as *Category 2*, see Fig. 2d. Because of the approximately flat reliability line, there is no relationship between the forecast probabilities and the frequencies of the observed event; the forecast system is not useful for decision making. An example from System 4 for Category 2 forecasts are the predictions of dry summers (JJA) over Southern Europe (Fig. 3d).

The poorest category of forecast reliability, *Category 1*, summarises forecasts where the reliability line has a negative slope implying an inverse relation between the forecast probabilities and the frequencies of the observed event (Fig. 2e). These forecasts are dangerously useless for decision making because they not only provide no useful information but also can mislead the users of the forecasts with unreliable information. Dry summer (JJA) forecasts for Northern Europe from System 4 fall into this very unreliable category, see Fig. 3e.

In principle, the raw probabilistic output from such seasonal forecast systems can be calibrated empirically using a training sample of data (*Wilks, 2011*). However, with limited training data (30 years is not a large sample), such empirical calibration cannot be assumed to produce reliable out-of-sample probability forecasts. As such, a key aspiration of any operational forecast centre must be to produce reliable forecasts without recourse to empirical calibration.

## 5. Reliability of System 4

This Section gives a summary of how reliable the System 4 seasonal forecasts for near-surface temperature and precipitation in JJA and DJF are in terms of our 5 categories of reliability as defined in Section 4.

In Figure 4 we show the reliability categories for 2m temperature in DJF and JJA over 21 land regions around the world. Almost all of the areas fall into the first three categories from perfect to marginally useful reliability. Only the Northern Asia region for cold DJF events has been classified as not useful (Category 2).

Remarkably, there are a number of extra-tropical regions where reliability is found to be perfect (Category 5): the east and west coasts of North America and parts of China and East Asia in DJF, and South America, Southern Africa and Australia in austral winter in JJA.

For Europe, all winter predictions of temperature fall into the marginally useful Category 3, whereas in summer the temperature forecasts over Europe is improved. Cold anomalies over Northern Europe are classified to have perfect reliability. Category 4 reliability of still being potentially very useful after calibration has been found for Southern Europe and Category 3 performance is shown for warm summers over Northern Europe.

Over the extended tropical areas temperature forecasts in both seasons are classified as either Category 5 or Category 4, except for cold anomalies over western tropical Africa in JJA which have Category 3 reliability. The Sahara region is an area that consistently falls into Category 5 for cold and warm temperature events in JJA and DJF.

Results for the reliability categorisation of precipitation forecasts are shown in Fig. 5 for wet and dry DJF and JJA seasons. Overall, the reliability performance for precipitation is poorer than for temperature with more regions being classified with lower categories. The marginally useful Category 3 is the most frequent category for precipitation forecasts in both seasons and for both wet and dry events.

Even though the overall performance is not very reliable, there are areas and events that are classified as perfectly or usefully reliable. Consistent regions among the events and seasons are Central America, Northern parts of South America and South-East Asia.

Over Northern Europe, the reliability of precipitation forecasts for winters (DJF) is not useful (Category 2) for dry events and marginally useful for wet events. Southern Europe falls into the middle Category 3 in winter and for the wet summer events. The reliability for dry summers over Europe is notably poor with Southern Europe classified as not useful (Category 2) and Northern Europe as dangerously useless (Category 1). Parts of Northern America fall into that lowest category, too.

To summarise these findings, Fig. 6 shows how many regions there are in each of the 5 reliability categories when accumulated over all seasons and tercile events. The most frequent category for the temperature forecasts is Category 4 describing forecasts with good reliability that can still be useful for decision making. The perfect reliability Category 5 has also been found for a lot of regions while only one of the regions fell into the not useful Category 2. None of the temperature forecasts was classified as dangerously useless in terms of its reliability.

As mentioned above, the most frequent reliability category for precipitation is the marginally useful Category 3. Such forecasts are not very reliable but might potentially be marginally useful for some applications. The category with the second highest number of regions is the one of perfect reliability, which is an optimistic result for the usefulness of seasonal forecasts of precipitation. However, there are substantially more cases of areas that have poor reliability (Categories 1 and 2); users should not use these forecasts for decision making in these regions as they can be dangerously misleading.

## 6. How can seasonal forecast reliability be improved?

The results above suggest that we still have some way to go before it can be said that the goal of providing users with reliable forecasts has been achieved, particularly for precipitation and away from the El Niño region. There can be little doubt that the ability to represent physical processes accurately is key to improved reliability. In a recent study based on integrations within the project Athena (*Jung et al., 2012*), *Dawson et al. (2012)* were able to show in AMIP integrations that the ECMWF model could simulate the non-Gaussian structure of observed Euro-Atlantic weather regimes more accurately in a model with spectral resolution T1279 (approx. 15 km) than with resolution T159 (approx. 125 km). It is plausible that the improved simulation of such weather regimes in a T1279 model is associated with better representation of topography on the one hand, and with a more realistic representation of Rossby wave breaking on the other. Improved simulation of stratospheric processes through finer vertical resolution is also expected to impact seasonal forecasts (*Cagnazzo and Manzini, 2009*; *Marshall and Scaife, 2009*).

A fundamental question, but one that is probably unanswerable until the tools are available to answer it, is whether perfect forecast reliability can only be achieved when convective cloud systems (with scales of just a few kilometres) are resolved explicitly. Much of the skill of seasonal forecasts originates in the tropics, and moist convection is a dominant form of instability in the tropics.

Seasonal forecasting with such cloud-resolved models will require exascale computing capability.

A better representation of other Earth System components is also likely to improve reliability. For example, *Weisheimer et al. (2011a)* showed that a better representation of land surface processes led to remarkably good probabilistic forecast of the summer 2003 heat wave.

On the other hand, since the climate system is chaotic, it is necessary to represent inevitable uncertainties in the representation of processes which have to be parameterised. A programme to represent parameterisation uncertainty has been on-going for some time at ECMWF (*Buizza et al., 1999; Palmer, 2001; Palmer, 2012*), and on the monthly and seasonal timescales there is evidence that it is competitive with, and for temperature predictions can outperform, the more standard multi-model ensemble approaches to the representation of model uncertainty (*Weisheimer et al., 2011b*).

There can be little doubt that the value to society of reliable non-climatological predictions of seasonal climate. However, to develop a high resolution system with accurate stochastic representations of model uncertainty in all relevant components of the Earth System, is not only a formidable technical challenge, it may be one that will require computing resources which are unavailable to individual institutes in the foreseeable future. A possible route to achieve the goal of a reliable seasonal climate prediction system, based on much stronger international collaboration, has been presented elsewhere (*Shukla et al., 2010; Shapiro et al., 2010; Palmer, 2011; Palmer, 2012*).

## 7. Conclusions

Let us return to the question posed in the Introduction. What constitutes a "5", to which a seasonal forecast system should aspire? Here we propose the following broad criterion for rating a seasonal forecast system a "5": when the system predicts probabilities $\rho(X)$ that are substantially different from the climatological distribution $\rho_C(X)$ then these probabilities can be relied on, and acted on by decision makers. Note that we make no firm statement about how often such situations arise. It may be that in the majority of cases the forecast system does not predict probabilities that differ substantially from $\rho_C(X)$. If this is the case, then the probabilistic skill score may not differ substantially from zero. However, for such a forecast system, a user will not make a bad decision based on unreliable forecast information.

Reliability of seasonal forecasts can also be considered relevant in the context of seamless prediction of weather and climate: the reliability of climate predictions on the seasonal time scale can provide constraints for the trustworthiness of climate change projections. Reliability diagrams provide a means to calibrate climate change probabilities and discount these climate change probabilities if the seasonal forecasts can be shown to not be reliable (*Palmer et al., 2008*).

The ECMWF seasonal forecast System 4 cannot be rated a "5" for all regions of the world, and for all variables. We have shown that for surface temperature, and even more for precipitation, forecast probabilities are not reliable when different from climatology and away from the El Niño region. Based on current performance our current capability to forecast seasonal climate was rated between 1 and 5 depending on variable and scale. However, given expected increases in resolution, and better stochastic representations of model uncertainty, we see no reason why this should not rise to 5 overall in the coming 30 years.

## 8. Acknowledgements

This study was supported by the EU project SPECS funded by the European Commission's Seventh Framework Research Programme under the grant agreement 308378. We thank Liz Stephens, Susanna Corti, Sophie Haines and David MacLeod for helpful comments on the draft of this manuscript.

**List of figures**

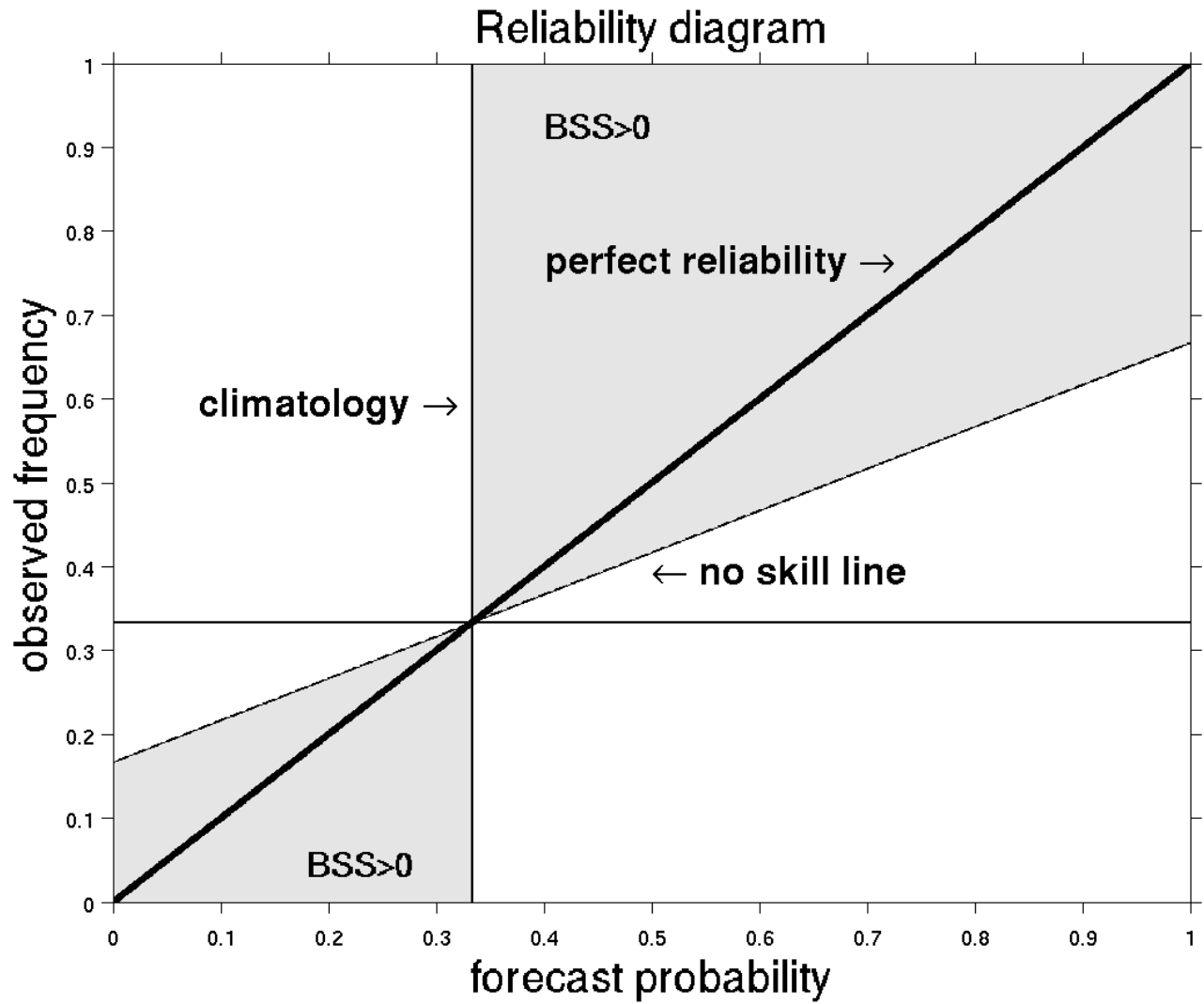

**Figure 1:** What is a reliability diagram? A reliability diagram shows the observed frequencies of an event as a function of its forecast probability. The thick diagonal line indicates perfect reliability. The thin horizontal and vertical lines show the climatological probabilities of the event in the forecasts and observations (here 1/3 for tercile events). The dashed line is the no-skill line and defines areas in the diagram where data contribute positively to the Brier Skill Score (grey areas).

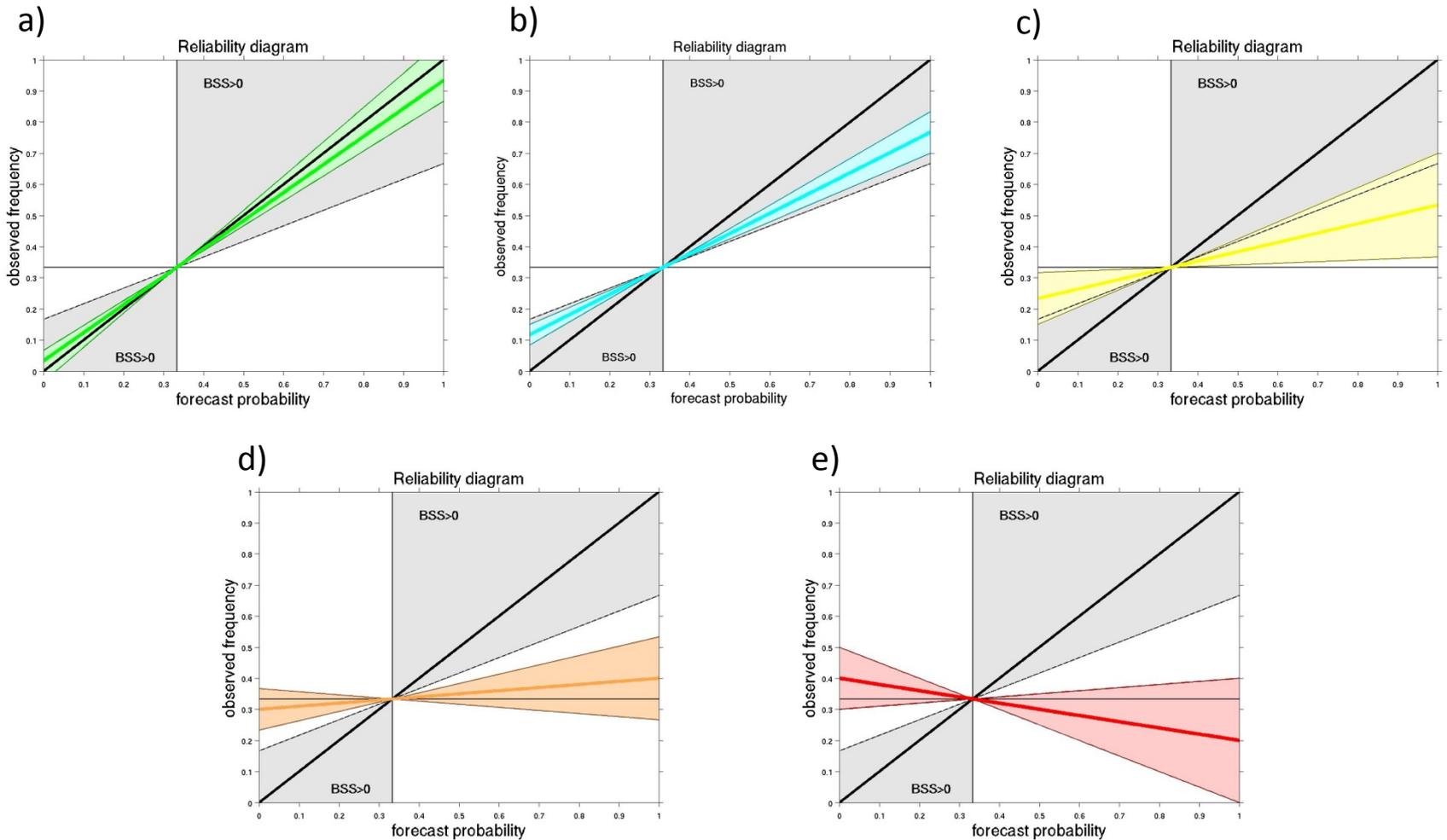

**Figure 2:** Five categories of reliability: a) 5 - perfect, b) 4 - after calibration still very useful for decision making, c) 3 - marginally useful after calibration, d) 2 - not useful, and e) 1 - dangerously useless.

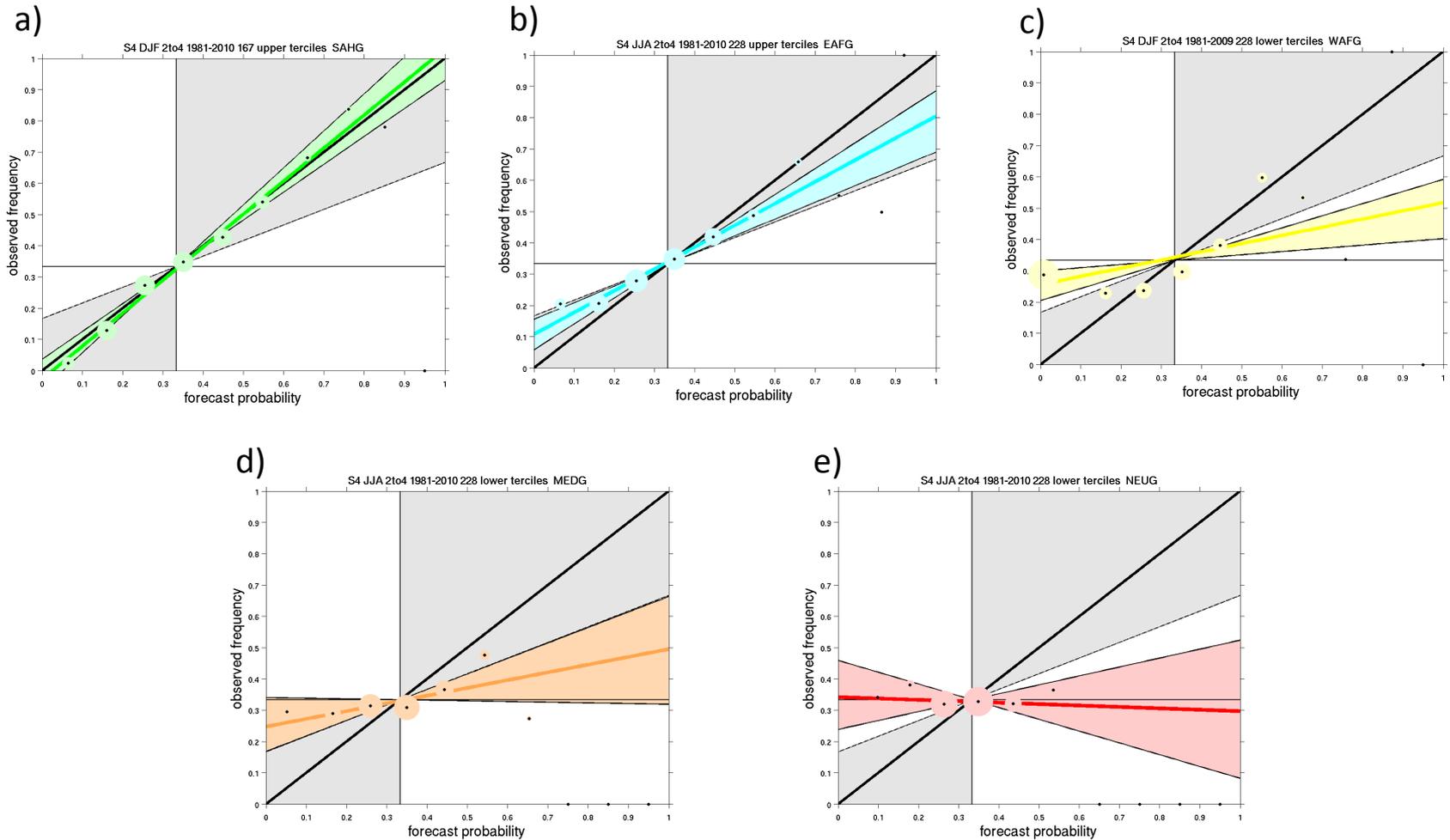

**Figure 3:** System 4 examples of the five categories of reliability: a) 5 - warm DJF over the Sahel, b) 4 - wet JJA over East Africa, c) 3 - dry DJF over West Africa, d) 2 - dry JJA over Southern Europe, and e) 1 - dry JJA over Northern Europe.

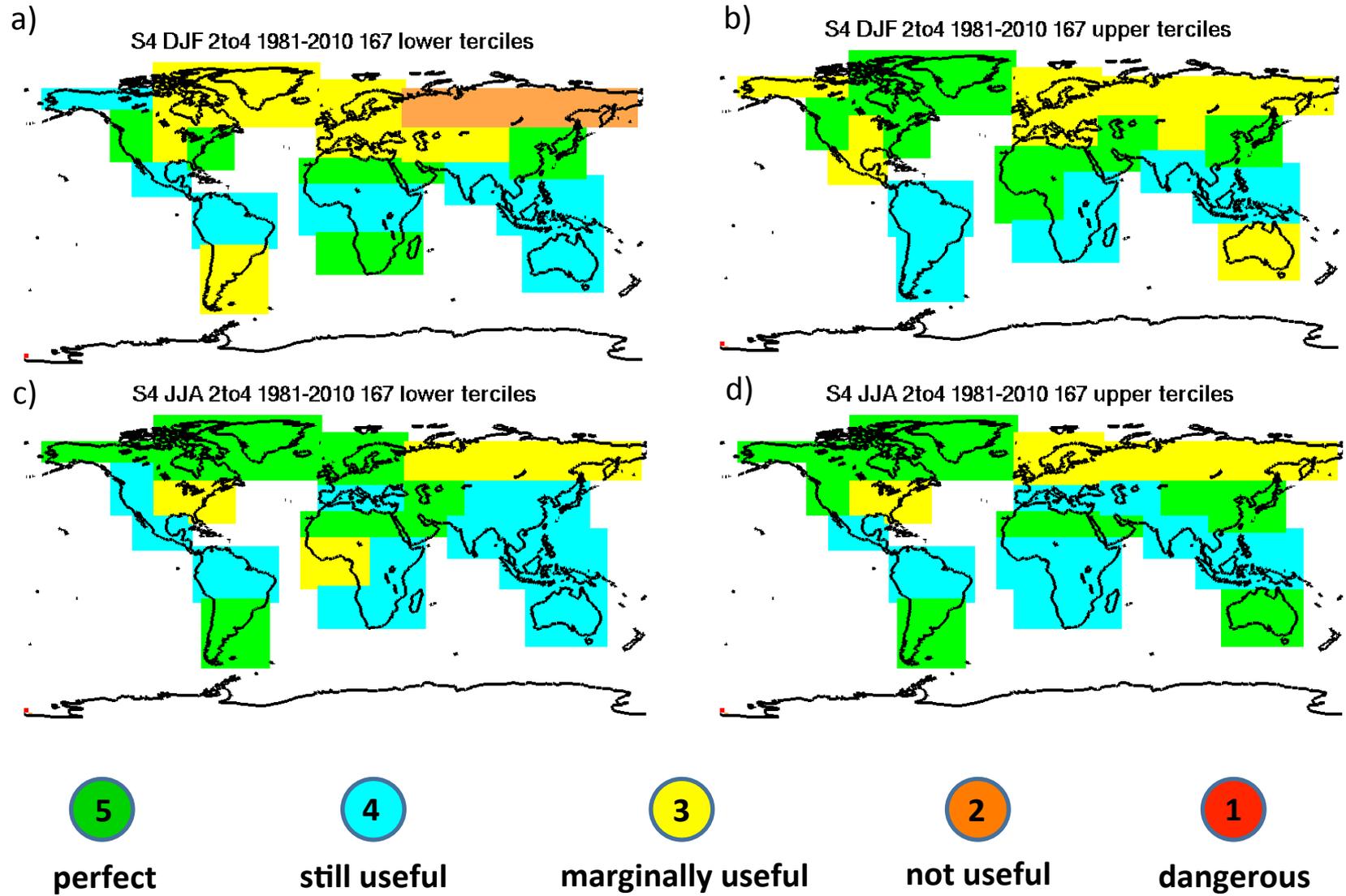

**Figure 4:** Reliability of System 4 seasonal forecasts for 2m temperature. a) cold DJF, b) warm DJF, c) cold JJA, and d) warm JJA.

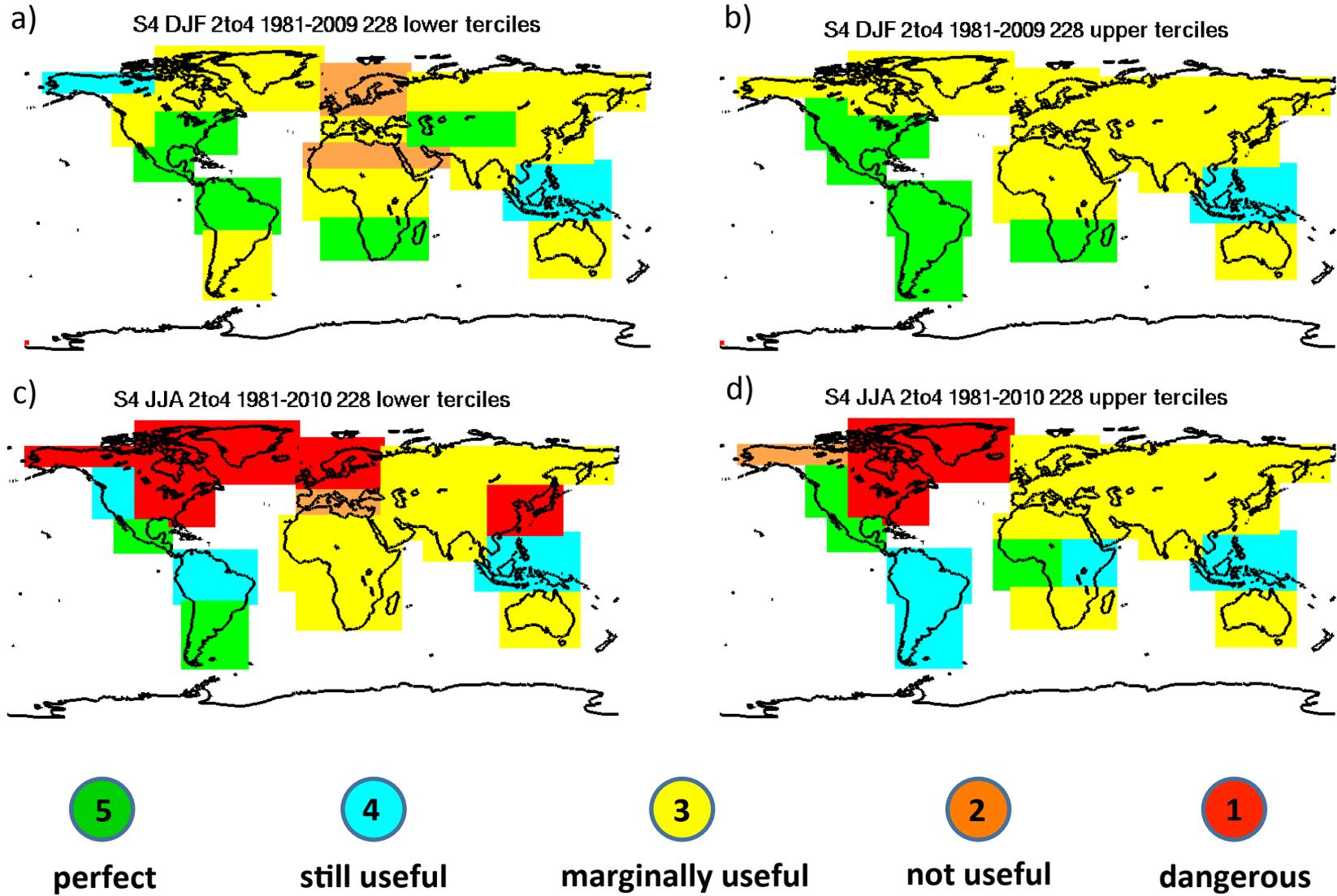

**Figure 5:** Reliability of System 4 seasonal forecasts for precipitation. a) dry DJF, b) wet DJF, c) dry JJA, and d) wet JJA.

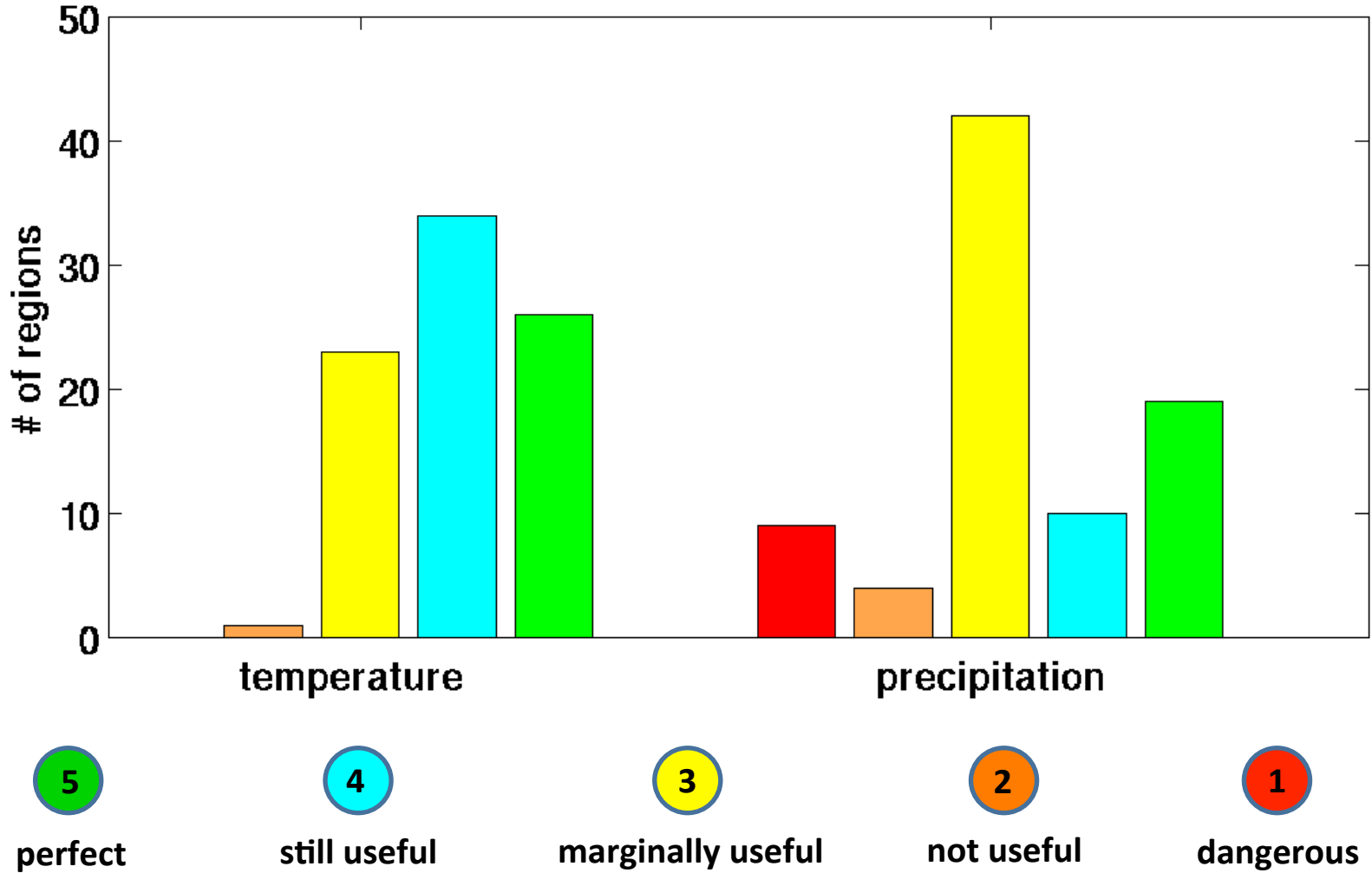

**Figure 6:** Number of regions that fall into each reliability category summed over all four events for temperature and precipitation.